\documentclass[aps,twocolumn,preprintnumbers,amsmath,amssymb,nofootinbib]{revtex4}
\usepackage[dvips]{color}
\usepackage{graphicx}
\usepackage{dcolumn}
\usepackage{bm}
\usepackage{xcolor}

\usepackage{amsmath,amssymb}
\usepackage{comment}

\usepackage[colorlinks]{hyperref}
\hypersetup{
linkcolor = blue,
citecolor = blue,
urlcolor = black
}

\begin{document}

\title{Perfect transmission of Higgs modes via antibound states}

\author{Takeru Nakayama$^1$}\email{t.nakayama@issp.u-tokyo.ac.jp}
\author{Shunji Tsuchiya$^2$}\email{tsuchiya@phys.chuo-u.ac.jp}
\affiliation{$^1$Institute for Solid State Physics, University of Tokyo, 
5-1-5 Kashiwanoha, Kashiwa, Chiba 277-8581, Japan}
\affiliation{$^2$Department of Physics, Chuo University,
1-13-27 Kasuga, Bunkyo-ku, Tokyo 112-8551, Japan}

\date{\today}

\begin{abstract} 
We study tunneling properties of Higgs modes in superfluid Bose gases in optical lattices in the presence of a potential barrier introduced by local modulation of hopping amplitude. Solving the time-dependent Ginzburg-Landau equation, Higgs modes are found to exhibit perfect transmission through a potential barrier if the barrier strength is weak. There exists, on the other hand, localized Higgs bound states in the presence of a strong potential barrier. We find that the perfect transmission disappears at the critical barrier strength above which one of the odd antibound state turns into a true bound state. We demonstrate that the perfect transmission of Higgs modes is mediated by resonance with the antibound states of Higgs modes.
\end{abstract}

\maketitle
\section{Introduction}

Spontaneous symmetry breaking is a central concept in condensed matter physics. Two types of collective mode emerge in association with spontaneous breaking of continuous symmetries. One is the Nambu-Goldstone (NG) mode \cite{nambu-60,goldstone-61} and the other is the Higgs mode \cite{higgs-64,littlewood-81}. The NG mode is a gapless excitation that arises from phase fluctuation of the order parameter. Nambu-Goldstone modes dominate low-energy properties of the system and have been studied in various condensed matter systems.
On the other hand, the Higgs mode is a gapped mode that involves amplitude fluctuation of the order parameter. Since it is difficult to excite and probe Higgs modes selectively, it is only recently that experimental progress has enabled systematic investigation of Higgs modes in condensed matter systems \cite{pekker-15}.\footnote[1]{For recent progress in the study of Higgs modes in condensed matter systems, see, for example,  Ref.~\cite{pekker-15} }
 In particular, Bose superfluids in optical lattices offer an ideal playground for investigating various aspects of Higgs modes due to the high controllability of the system \cite{bissbort-11,endres-12}.
\par
The tunnel effect is a pure quantum-mechanical phenomenon and has attracted much interest. 
Collective modes exhibit interesting tunneling properties that are very different from those of single particles. For example, NG modes in Bose-Einstein condensates (BECs) have been predicted to perfectly transmit a potential barrier in the low-energy limit, which is referred to as anomalous tunneling \cite{kagan-03,danshita-06,kato-08,tsuchiya-09,kato-12}. It has been found that NG modes in Bose superfluids in optical lattices cause Fano resonance mediated by Higgs bound states when they tunnel through potential barriers \cite{nakayama-15}. However, little is known about tunneling properties of Higgs modes.
\par
In the present paper, extending our recent work \cite{nakayama-15}, we study tunneling of Higgs modes in Bose superfluids in optical lattices. Solving the time-dependent Ginzburg-Landau (TDGL) equation that describes the superfluid dynamics in the vicinity of the phase boundary to the Mott insulating state, we show that Higgs modes perfectly transmit a potential barrier introduced by local modulation of the hopping amplitude when the barrier potential is weak. The perfect transmission does not occur for a strong potential barrier when the odd bound state of Higgs modes exist. We investigate the origin of the perfect transmission and find that it is mediated by the antibound states of Higgs modes.  
\par
This paper is organized as follows.
In Sec.~\ref{sec:Model} we introduce the Bose-Hubbard (BH) model and the TDGL equation including the effects of external potentials.
In Sec.~\ref{sec:Higgs_tunneling} we study the tunneling problem of Higgs modes solving the TDGL equation in the presence of a $\delta$-function potential and a rectangular potential. We show that perfect transmission of Higgs modes occurs and discuss the origin of it relating it to the antibound states.   
In Sec.~\ref{sec:Conclusion} the results are summarized.

\section{Model}\label{sec:Model}

We consider bosons trapped in a cubic optical lattice. They are well described by the tight-binding BH model \cite{fisher-89,jaksch-98}
\begin{eqnarray}
\mathcal H=
-\sum_{{\bm i},{\bm j}}J_{{\bm i},{\bm j}}b_{\bm i}^\dagger b_{\bm j}
-\sum_{\bm i} \mu_{\bm i} b_{\bm i}^\dagger b_{\bm i}
+\frac{U}{2}\sum_{\bm i} b_{\bm i}^\dagger b_{\bm i}^\dagger b_{\bm i} b_{\bm i}.
\label{eq:BH}
\end{eqnarray}
The vector ${\bm i}\equiv \sum_{\alpha=1}^{d}i_{\alpha}{\bm e}_{\alpha}$ denotes the lattice site, where $i_{\alpha}$ is an integer, $d$ is the spatial dimension, and ${\bm e}_{\alpha}$ is a unit vector in the direction $\alpha$. In addition,
$b_{\bm i}^\dagger$ ($b_{\bm i}$) is a creation (annihilation) operator of bosons at site ${\bm i}$, and $U>0$ is  the on-site repulsive interaction. The chemical potential $\mu_{\bm i}\equiv \mu_0 - V_{\bm i}$ includes the homogeneous contribution $\mu_0$ and the external potential $V_{\bm i}$. Further, $J_{{\bm i},{\bm j}}=\sum_{\alpha}(J^{(\alpha)}_{\bm j} \delta_{{\bm i},{\bm j}+{\bm e}_{\alpha}}+ J^{(\alpha)}_{{\bm j}-{\bm e}_{\alpha}} \delta_{{\bm i},{\bm j}-{\bm e}_{\alpha}})$ is the hopping matrix element between adjacent sites, where $J^{(\alpha)}_{\bm j}$ denotes the hopping amplitude between sites ${\bm j}$ and ${\bm j}+{\bm e}_{\alpha}$.
We neglect the harmonic trapping potential for simplicity. We set $\hbar=1$ and assume zero temperature throughout the paper.
\par
In previous work \cite{nakayama-15} we proposed to study tunneling effects of the NG mode in the superfluid phase by introducing the local shift of the chemical potential $V_{\bm i}$ and hopping amplitude $J_{\bm i}^{(\alpha)}$ independently. The former can be introduced by imposing an optical dipole potential, while the latter can be introduced by imposing an additional lattice potential in the Gaussian profile with the same lattice spacing as that of the overall potential (see Fig.~3 in Ref.~\cite{nakayama-15}). A local potential barrier that modulates the hopping amplitude locally can be created also by using a digital micromirror device \cite{islam-15}. In this paper, we focus on the latter for simplicity and set $V_{\bm i}=0$.
We further assume the inhomogeneity of the hopping only in the $x$ direction: $J_{\bm i}^{(\alpha)}= J+J'_{i_1}\delta_{\alpha,1}$. The system is assumed to be homogeneous in the $y$ and $z$ directions.
\par
The TDGL equation that governs the dynamics of the superfluid order parameter $\psi(\bm x,t)$ can be derived in the vicinity of the superfluid -- Mott-insulator (SF-MI) transition point by taking the low-energy and continuum limit~\cite{fisher-89,sachdev-11}. The TDGL equation including the effects of the inhomogeneous hopping reads \cite{nakayama-15}
\begin{eqnarray}
iK_0\frac{\partial \psi}{\partial t} \!-\! W_0\frac{\partial^2 \psi}{\partial t^2}
\!=\!
\left(-\frac{\nabla^2}{2m^{\ast}}+r_0+v_r+u_0|\psi|^2\right)\psi,
\label{eq:tdgl_b}
\end{eqnarray}
where $K_0$, $W_0$, $r_0 (<0)$, $m^{\ast}$, and $u_0$ are functions of the original parameters in the BH model $(J, \mu_0, U)$ (their expressions are given in Appendix A in Ref.~\cite{nakayama-15}). Here,  $v_{r}(x)\equiv - 2J'(x)$ represents the potential due to the inhomogeneous hopping in the continuum limit $J'_{i}\to J'(x)$. 
\par
We assume a commensurate filling, which results in the approximate particle-hole symmetry in the vicinity of the SF-MI transition point \cite{altman-02,huber-07,huber-08}. Since the TDGL equation should be invariant under the charge-conjugation transformation $\psi \leftrightarrow \psi^{\ast}$ in the presence of the particle-hole symmetry, the first-order time-derivative term should vanish $K_0=0$. Thus, Eq.~(\ref{eq:tdgl_b}) reduces to the nonlinear Klein-Gordon equation in the relativistic field theory~\cite{higgs-64} that exhibits the emergent Lorentz invariance.
\par
We employ the TDGL equation in the dimensionless form,
\begin{equation}
-\frac{\partial^2\tilde{\psi}}{\partial \tilde{t}^2}=\left(-\frac{\tilde{\nabla}^2}{2}-1+|\tilde{\psi}|^2+\tilde{v}_r\right)\tilde{\psi},
\end{equation}
where the variables are normalized as
\begin{eqnarray}
\begin{split}
\tilde{\psi}=\psi/(\left|r_0\right|/u_0)^{1/2}, \quad \tilde{t}=t(\left|r_0\right|/W_0)^{1/2},\\
\tilde{\bm{x}}=\bm{x}/\xi,\quad \tilde{v}_r=v_r/\left|r_0\right|, 
\end{split}
\label{eq:Dless}
\end{eqnarray}
with $\xi \equiv (m_{\ast}\left|r_0\right|)^{-1/2}$ denotes the healing length. Hereafter, we omit the tilde.
\par
We consider fluctuations of the order parameter $\psi({\bm x},t)$ around its static solution $\psi_0({\bm x})$,
\begin{eqnarray}
\psi({\bm x},t)=\psi_0({\bm x})+\mathcal{U}({\bm x})e^{-i\omega t}+\mathcal{V}({\bm x})^*e^{i\omega t}.
\end{eqnarray}
Here $S(\bm{x})\equiv\mathcal{U}(\bm{x})-\mathcal{V}(\bm{x})\propto \delta \theta(\bm{x})$ and $T(\bm{x})\equiv \mathcal{U}(\bm{x})+\mathcal{V}(\bm{x})\propto \delta n(\bm{x})$ represent phase and amplitude fluctuations of the order parameter, respectively.
In addition, $\psi_0({\bm x})$ satisfies the static Gross-Pitaevskii (GP) equation~\cite{pitaevskii-61}:
\begin{eqnarray}
 \left(-\frac{\nabla^2}{2}-1+|\psi_0({\bm x})|^2+v_r(x)\right)\psi_0(\bm{x})=0~.
\label{eq:static_3d}
\end{eqnarray}
The equations for phase and amplitude fluctuations read, respectively,
\begin{eqnarray}
\begin{split}
\left(-\frac{\nabla^2}{2}-1+|\psi_0({\bm x})|^2+v_r(x)\right)S(\bm{x})
=\omega^2S(\bm{x})~, \label{eq:S(x)_3d}
\end{split}
\\
\begin{split}
\left(-\frac{\nabla^2}{2}-1+3|\psi_0({\bm x})|^2+v_r(x)\right)T(\bm{x})
=\omega^2T(\bm{x})~. \label{eq:T(x)_3d}
\end{split}
\end{eqnarray}
Equations~(\ref{eq:S(x)_3d}) and (\ref{eq:T(x)_3d}) demonstrate that phase and amplitude fluctuations are uncoupled due to the particle-hole symmetry \cite{tsuchiya-18}.
\par
Without the potential barrier ($v_r=0$), assuming plane wave solutions $(S(\bm{x}),T(\bm{x}))=(S_{\bm{k}},T_{\bm{k}})e^{i\bm{k}\cdot\bm{x}}$, we obtain the dispersion relations for the NG and Higgs modes as, respectively, 
\begin{eqnarray}
\begin{split}
&\quad \omega^2=\frac{k^2}{2}, \label{NGdisp}\\
&\quad \omega^2=\frac{k^2}{2}+\Delta^2. \label{Higgsdisp}
\end{split}
\end{eqnarray}
The NG mode indeed has a gapless dispersion, while the Higgs mode has the energy gap $\Delta=\sqrt{2}$. Since phase and amplitude are uncoupled, the NG and Higgs modes involve pure phase and amplitude oscillations, respectively.

\section{Tunneling problem of Higgs modes}\label{sec:Higgs_tunneling}
We study tunneling problem of Higgs modes through a potential barrier $v_r(x)$. 
Since the static order parameter $\psi_0(x)$ is assumed to be homogeneous in the $y$ and $z$ directions, the GP equation reduces to
\begin{eqnarray}
 \left[-\frac{1}{2}\frac{d^2}{dx^2}-1+|\psi_0(x)|^2+v_r(x)\right]\psi_0(x)=0~.
\label{eq:static}
\end{eqnarray}
We assume the plane wave forms in the $y$ and $z$ directions as
\begin{eqnarray}
S({\bm x})=S_{\rm 1D}(x)e^{i\bm{k}_{\parallel}\cdot \bm{x}_{\parallel}},
\\
T({\bm x})=T_{\rm 1D}(x)e^{i\bm{k}_{\parallel}\cdot \bm{x}_{\parallel}},
\end{eqnarray}
where $\bm{k}_{\parallel}=(k_y,k_z)$ and $\bm{x}_{\parallel}=(y,z)$.
In the following analysis, we assume that the NG and Higgs modes propagate only in the $x$ direction,
 i.e., $\bm{k}_{\parallel}=\bm{0}$. Thus, Eqs.~(\ref{eq:S(x)_3d}) and (\ref{eq:T(x)_3d}) reduce to
\begin{eqnarray}
\left[-\frac{1}{2}\frac{d^2}{dx^2}-1+|\psi_0(x)|^2+v_r(x)\right]S_{\rm 1D}(x)=\omega^2S_{\rm 1D}(x)~, \label{eq:S(x)}\\
\left[-\frac{1}{2}\frac{d^2}{dx^2}-1+3|\psi_0(x)|^2+v_r(x)\right]T_{\rm 1D}(x)=\omega^2T_{\rm 1D}(x)~. \label{eq:T(x)}
\end{eqnarray}
We henceforth denote $S_{\rm 1D}(x)$ and $T_{\rm 1D}(x)$ by $S(x)$ and $T(x)$ for brevity.

\subsection{$\delta$-function potential barrier}
We first study tunneling of Higgs modes across a $\delta$-function potential barrier $v_r(x)=V_r\delta(x)$ ($V_r>0$). Note that any potential barrier that spatially varies in the order of the lattice spacing can be approximated as a $\delta$-function potential in the vicinity of the phase boundary with the MI phase, where the healing length $\xi$ gets much larger than the lattice spacing. The analytic solution of Eq.~(\ref{eq:static}) under the $\delta$-function potential \cite{kovrizhin-01} is given by 
\begin{equation}
\psi_0(x)=\tanh(|x|+x_0),
\label{eq:static_kink}
\end{equation}
where $x_0$ is determined by the boundary conditions at $x=0$,
\begin{eqnarray}
&\displaystyle \psi_0(-0)=\psi_0(+0), \label{bdcondp1}\\
&\displaystyle \left.\frac{d \psi_0}{d x}\right|_{+0}-\left.\frac{d \psi_0}{d
		x}\right|_{-0}=2V_r\psi_0(0).
\label{bdcondp2}
\end{eqnarray}
We thus obtain
\begin{equation}
\tanh(x_0)=-\frac{V_r}{2}+\sqrt{\frac{V_r^2}{4}+1}\equiv \eta.
\end{equation}
Note that $\eta=\psi_0(0)$, which is the amplitude of the order parameter under the barrier, satisfies $0\leq \eta\leq 1$. Here $\eta$ decreases from $\eta(V_r=0)=1$ with increasing $V_r$ and has the asymptotic form $\eta\sim 1/V_r$ as $V_r\gg 1$.
\par
We consider tunneling of Higgs modes through the $\delta$-function potential barrier. We assume that Higgs modes with energy $E\geq\Delta$ and wave vector $k=\sqrt{2}\sqrt{E^2-\Delta^2}$ are injected from $x\to-\infty$. 
The solution of Eq.~(\ref{eq:T(x)}) can be written in a linear combination of the plane waves on a static kink condensate \cite{nakayama-15,lamb} as
\begin{widetext}
\begin{eqnarray}
T(x)=\left\{
\begin{array}{ll}
   \dfrac{3\psi_0^2+3ik\psi_0-k^2-1}{2+3ik-k^2}e^{ikx}+r_{\rm{h}}\dfrac{3\psi_0^2-3ik\psi_0-k^2-1}{2-3ik-k^2}e^{-ikx} & (x<0) \\\\
    t_{\rm{h}} \dfrac{3\psi_0^2-3ik\psi_0-k^2-1}{2-3ik-k^2}e^{ikx} & (x>0)
  \end{array} \right. .\label{eq:higgs_tunnel_WF}
\end{eqnarray}
\end{widetext}
Here, $t_{\rm h}$ and $r_{\rm h}$ denote the transmission and reflection amplitudes, respectively.
The asymptotic form of Eq.~(\ref{eq:higgs_tunnel_WF}) far away from the potential barrier is given by
\begin{eqnarray}
T(x)\rightarrow
\left\{
\begin{array}{l}
e^{ikx} + r_{\rm h}e^{-ikx} \quad(x\to-\infty) \\
\\
t_{\rm h} e^{ikx} \quad(x\to\infty)
\end{array}
\right..
\label{eq:asymptoticT}
\end{eqnarray}
The reflection and transmission probabilities of Higgs modes are given by $\mathcal{R}\equiv |r_{\rm h}|^2$ and
$\mathcal{T}\equiv |t_{\rm h}|^2$, respectively. They satisfy the conservation law $\mathcal{R}+\mathcal{T}=1$.
\par
\begin{figure}[t]
\begin{center}
\includegraphics[width=\linewidth]{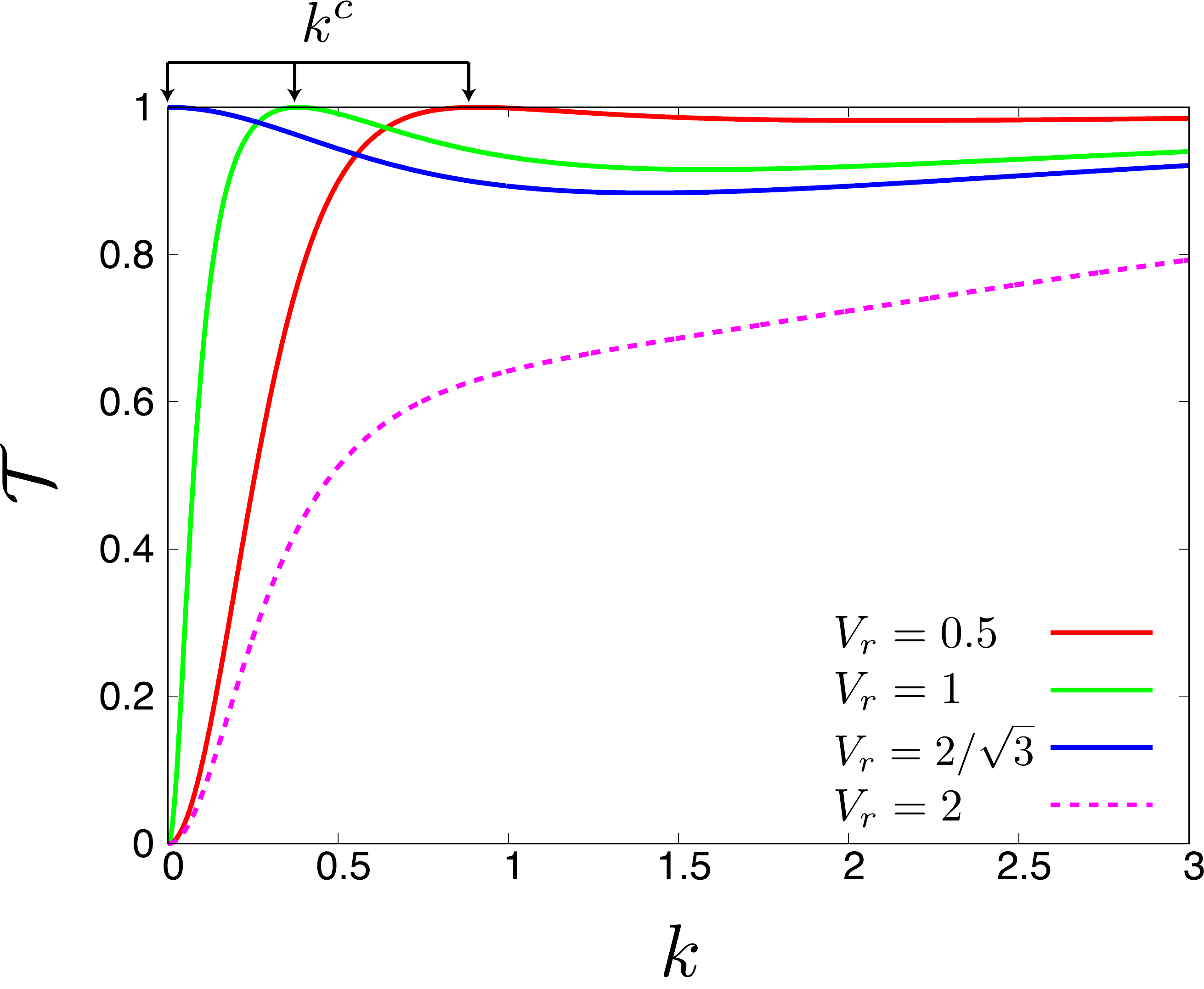}
\caption{Transmission probability of Higgs modes $\mathcal{T}$ through a $\delta$-function potential $V_r\delta(x)=V_r\delta(x)$ ($V_r>0$) as a function of the wave vector of the injected Higgs mode $k$ for various values of the barrier strength $V_r$. The horizontal axis is in the unit of $\xi^{-1}$.}
\label{fig:delta}
\end{center}
\end{figure}

The coefficients $r_{\rm h}$ and $t_{\rm h}$ are determined so as to satisfy the boundary conditions:
\begin{eqnarray}
&T(-0)=T(+0)\label{eq:higgs_connection1},\\
&\displaystyle \left.\frac{d T}{d x}\right|_{+0}-\left.\frac{d T}{d x}\right|_{-0}=2V_rT(0).\label{eq:higgs_connection2}
\end{eqnarray}
From the above conditions, the transmission amplitude of Higgs modes can be calculated as
\begin{eqnarray}
t_{\rm{h}}&=
&e^{i\delta}\frac{ik(k^2+1)(k^2+4)}{\left(c_1+V_rc_2\right)c_2},
\label{eq:t_amp}
\end{eqnarray}
where 
\begin{eqnarray}
e^{i\delta}&=&\left(2-3ik-k^2\right)/\left(2+3ik-k^2\right),\\
c_1&=&ik^3-3\eta k^2-ik(6\eta^2-4)+6\eta(\eta^2-1),\label{eq:c1}\\
c_2&=&-k^2-3i\eta  k+3\eta^2-1.\label{eq:c2}
\end{eqnarray} 
We thus obtain the transmission probability $\mathcal{T}(k)=|t_{\rm h}|^2$ as
\begin{eqnarray}
\mathcal{T}_{\rm{h}}^{-1}= 1+ V_r^2 \frac{(k^2+1-3\eta^2)^2(k^2+1+3\eta^2)^2}{k^2(1+k^2)^2(4+k^2)^2}.
\label{eq:higgs_transmission}
\end{eqnarray}
Equation~(\ref{eq:higgs_transmission}) shows that a perfect transmission ($\mathcal{T}=1$) occurs at $k^{\rm c}=\sqrt{3\eta^2-1}$, if the strength of the potential barrier is smaller than the critical value $V_r\leq 2/\sqrt{3}\equiv V_r^{\rm c}$ ($\eta\geq1/\sqrt{3}$).
Figure~\ref{fig:delta} shows the transmission probability (\ref{eq:higgs_transmission}) as a function of $k$ for various values of $V_r$. It exhibits the perfect transmission at $k^{\rm{c}}$ for weak potential barriers ($0<V_r\leq V_r^{\rm c}$). It is remarkable that the perfect transmission occurs in the long-wavelength limit $k\to0$ at the critical barrier strength $V_r^{\rm c}$. For strong potential barrier $V_r>V_r^{\rm c}$, perfect transmission no longer takes place.
\par
The origin of the perfect transmission is the main focus of our paper. One may naively think that the diminishing order parameter combined with the repulsive barrier constitutes an effective double-well potential for Higgs modes, so the perfect transmission is due to the resonant tunneling that is induced when the wavelength of Higgs modes matches the characteristic length of the double-well potential. 
However, this possibility is denied because the perfect transmission occurs even in the long wavelength limit $k\to0$.
The perfect transmission of Higgs modes reminds us of the anomalous tunneling of NG modes in BECs~\cite{kagan-03, danshita-06,kato-08,tsuchiya-09,kato-12}, where NG modes in BECs with momentum $k$ perfectly transmit a potential barrier in the limit $k\to 0$. The anomalous tunneling occurs because the wave function of the NG mode coincides with the condensate wave function at $k=0$. However, the wave function of Higgs modes $T(x)$ is not identical to the order parameter $\psi_0(x)$ at $k^{\rm c}$. Moreover, the perfect transmission does not occur for a sufficiently strong potential barrier $(V_r>V_r^{\rm c})$.

\subsection{Perfect transmission and antibound states}
We discuss the mechanism of the perfect transmission in more detail. In order to investigate the origin of the perfect transmission of Higgs modes, we study the solution of Eq.~(\ref{eq:T(x)}) in the following form~\cite{siegert-39}:
\begin{eqnarray}
T(x)=\left\{
\begin{array}{ll}
 A\dfrac{3\psi_0^2-3ik\psi_0-k^2-1}{2-3ik-k^2}e^{-ikx} & (x<0) \\\\
    B \dfrac{3\psi_0^2-3ik\psi_0-k^2-1}{2-3ik-k^2}e^{ikx} & (x>0)
  \end{array} \right. .
  \label{eq:Siegert}
\end{eqnarray}
The above form, which only involves the outgoing waves, is referred to as the Siegert condition~\cite{siegert-39,hatano-08}.
Under the boundary condition (\ref{eq:higgs_connection1}) and (\ref{eq:higgs_connection2}), we find that the solution satisfies one of the conditions
\begin{eqnarray}
c_1+V_rc_2&=&0 \label{eq:even_root},\\
c_2&=&0\label{eq:odd_root},
\end{eqnarray}
where $c_1$ and $c_2$ are given in Eqs.~(\ref{eq:c1}) and (\ref{eq:c2}).
\par
One obtains from Eq.~(\ref{eq:even_root}) an even-parity solution $A=B$ with the wave vector $k=i\kappa_{\rm e}$ ($\kappa_{\rm e}>0)$, where
\begin{widetext}
\begin{eqnarray}
6\kappa_{\rm e}&=&-2\left(V_r+3\eta\right)+\frac{2^{4/3}\left(V_r^2+6V_r\eta +3\right)}{\left(-2V_r^3+45V_r\eta^2+\sqrt{-4\left(V_r^2+6V_r\eta +3\right)^3+\left(2V_r^3-45V_r\eta^2\right)^2}\right)^{1/3}} \nonumber \\
&&+ 2^{2/3} \left(-2V_r^3+45V_r\eta^2+\sqrt{-4\left(V_r^2+6V_r\eta +3\right)^3+\left(2V_r^3-45V_r\eta^2\right)^2}\right)^{1/3}.
\end{eqnarray}
\end{widetext}
The excitation energy $E_{\rm e}=\sqrt{2-{\kappa_{\rm e}}^2/2}<\sqrt{2}$ is below the gap of the Higgs mode $\Delta=\sqrt{2}$. It is a true bound state involving an exponentially decaying wave function at $|x|\to\infty$. This solution is the even-parity Higgs bound state reported in Ref.~\cite{nakayama-15} which exists for any barrier strength $V_r>0$.
\par
On the other hand, solving Eq.~(\ref{eq:odd_root}), one obtains the odd-parity solutions with the wave vectors
\begin{eqnarray}
k_{\rm o}^\pm=\frac{i}{2}\left(\pm\sqrt{4-3\eta^2}- 3 \eta\right) \equiv i \kappa_{\rm o}^\pm.
\end{eqnarray}
Note that, since $4-3\eta^2>0$, $k_{\rm o}^\pm$ is pure imaginary. 
The solution involving exponentially decaying wave function $k_{\rm o}^+=i\kappa_{\rm o}^+$ ($\kappa_{\rm o}^+>0$) is a true bound state that exists when $V_r>V_r^{\rm c}$. Its binding energy is given by $E_{\rm o}=\sqrt{2-{\kappa_{\rm o}^+}^2/2}<\Delta$. This solution is the odd-parity Higgs bound state also reported in Ref.~\cite{nakayama-15}. 
The other solution $k_{\rm o}^-=i\kappa_{\rm o}^-$ ($\kappa_{\rm o}^-<0$) has an exponentially growing wave function at $|x|\to\infty$. Such a state, referred to as an antibound state~\cite{sasada-11,klaiman-10}, is not a true bound state. However, as we discuss below, it plays a crucial role in the perfect transmission of Higgs modes.
We note that, since Eq.~(\ref{eq:even_root}) is a cubic equation of $k$, there are two other even-parity solutions with complex $k$. These solutions are referred to as a resonant state if ${\rm Re}(k)>0$ and an antiresonant state if ${\rm Re}(k)<0$ \cite{sasada-11,klaiman-10}. However, it turns out that they are not related to the perfect transmission of Higgs modes.
\par
The odd-parity Higgs bound state becomes an antibound state with $\kappa_{\rm o}^+<0$ for $V_r<V_r^{\rm c}$. Meanwhile, the transmission probability $\mathcal T_{\rm h}$ exhibits the perfect transmission when $V_r<V_r^{\rm c}$.
Therefore, it is natural to suppose that the emergence of the perfect transmission may be related to the fact that the odd-parity Higgs bound state changes into an antibound state. We show that they are indeed closely related.
\par
\begin{figure}
\begin{center}
\includegraphics[width=\linewidth]{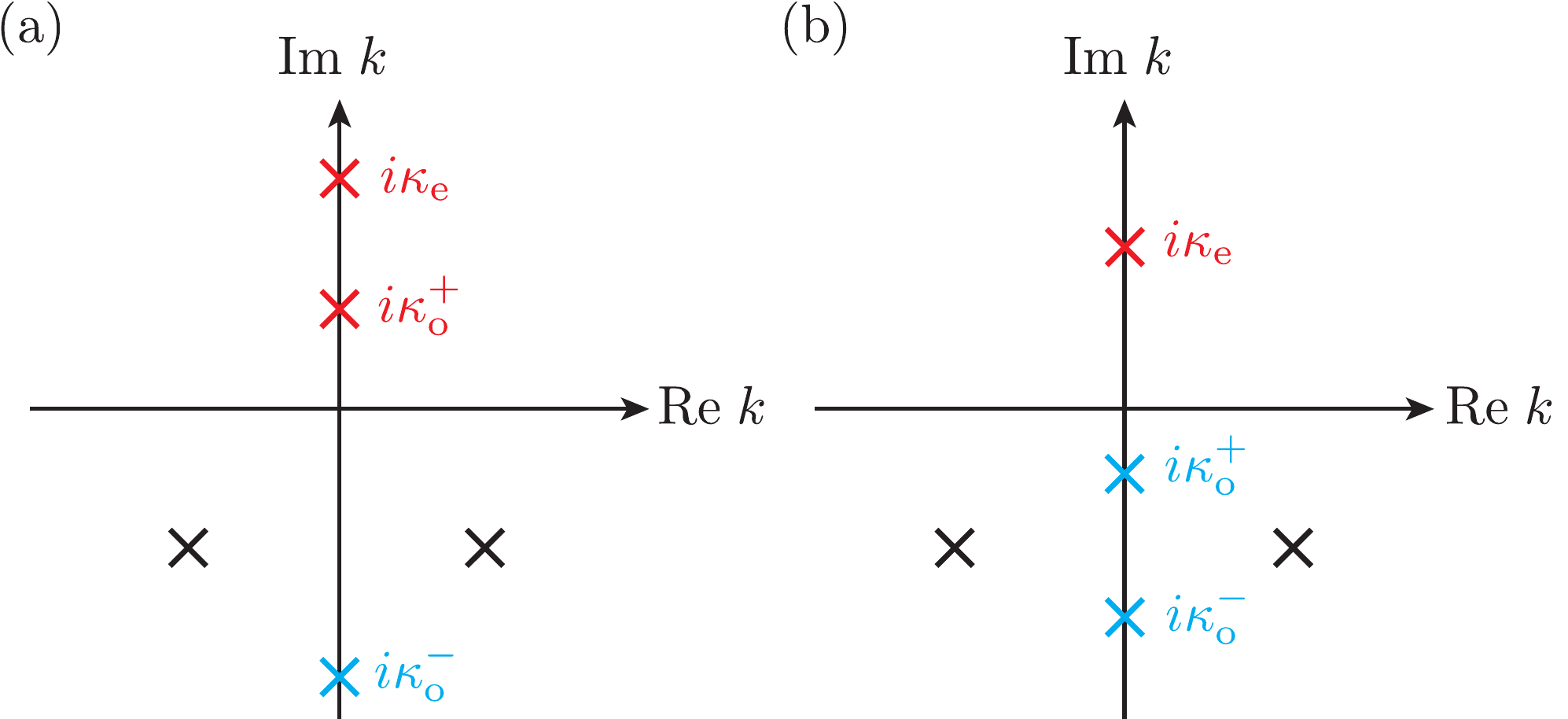}
\caption{Distribution of the poles of the transmission amplitude~(\ref{eq:t_amp}) on the complex $k$ plane when (a) $V_r >V_r^{\rm c}$ and (b) $V_r <V_r^{\rm c}$. The poles of the bound states and antibound states are located in the upper and lower half plane on the imaginary axis, respectively. The complex poles in the lower half plane correspond to the (anti)resonant states.}
\label{fig:comlex_k_plane}
\end{center}
\end{figure}

All the poles of the transmission amplitude~(\ref{eq:t_amp}) are given by the solutions of Eqs.~(\ref{eq:even_root}) and (\ref{eq:odd_root}). 
Figures~\ref{fig:comlex_k_plane}(a) and \ref{fig:comlex_k_plane}(b) show the distribution of the poles on the complex $k$ plane. If $V_r>V_r^{\rm c}$ [see Fig.~\ref{fig:comlex_k_plane} (a)], the poles of the even and odd bound states, $k=i\kappa_{\rm e}$ and $k=i\kappa_{\rm o}^+$, are located on the upper plane on the imaginary axis, while the pole of the antibound state $k=i\kappa_{\rm o}^-$ is on the lower plane on the imaginary axis. The pole of the odd bound state $k=i\kappa_{\rm o}^+$ moves downward as $V_r$ decreases. It enters the lower plane when $V_r<V_r^{\rm c}$ and becomes an antibound state, while other poles do not cross the real axis, as shown in Fig.~\ref{fig:comlex_k_plane}(b).

We can understand the origin of the perfect transmission of Higgs modes by examining the poles of the odd (anti)bound states.
The perfect transmission of Higgs modes cannot be interpreted in the usual resonance tunneling picture where the transmission probability has the Breit-Wigner form: $\mathcal T = (\Gamma/2)^2/\{(E-\varepsilon)^2+(\Gamma/2)^2\}$. Here, $\varepsilon$ is given by the real part of the pole and $\Gamma$ is the twice the imaginary part of the pole. In fact, the transmission probability near the peak cannot be approximated in this form. Instead, Eq.~(\ref{eq:higgs_transmission}) around $k\simeq k^{\rm{c}}$ can be approximated as
\begin{eqnarray}
\mathcal{T} \simeq \frac{k^2 \left(\kappa_{\rm o}^+ + \kappa_{\rm o}^-\right)^2}{\left(k^2 - \kappa_{\rm o}^+ \kappa_{\rm o}^-\right)^2+k^2 \left(\kappa_{\rm o}^+ +\kappa_{\rm o}^-\right)^2},
\label{eq:Klaimanform}
\end{eqnarray}
where $\kappa_{\rm o}^+ + \kappa_{\rm o}^-= -3\eta$ and $ \kappa_{\rm o}^+ \kappa_{\rm o}^-=3\eta^2-1$.
It is remarkable that the above equation coincides with the asymptotic form of the transmission probability for the double barrier potential in the presence of two antibound poles [see Eq.~(3) in Ref.~\cite{klaiman-10}].
Equation~(\ref{eq:Klaimanform}) shows that the position of the peak for the perfect transmission is determined by the product of $\kappa_{\rm o}^+$ and $\kappa_{\rm o}^-$.
If $V_r\leq V_r^{\rm c}$, the presence of the two odd antibound states with $\kappa_{\rm o}^\pm<0$ leads to the perfect transmission at $k^{\rm c}=\sqrt{\kappa_{\rm o}^+ \kappa_{\rm o}^-}$ in Eq.~(\ref{eq:Klaimanform}). The perfect transmission can thus be understood as being mediated by the antibound states. On the other hand, if $V_r>V_r^{\rm c}$, since one of the antibound states transforms into a true bound state, $\kappa_{\rm o}^+ \kappa_{\rm o}^-$ becomes negative and thus perfect transmission no longer occurs. At the critical barrier strength $V_r=V_r^{\rm c}$, the perfect transmission occurs precisely at $k^{\rm c}=0$.

\subsection{Rectangular potential barrier}

We demonstrate that the perfect transmission of Higgs modes is not due to the special feature of the $\delta$-function potential. For this purpose, we show that the perfect transmission also occurs in the presence of a rectangular potential barrier. We assume that Higgs modes are incident to a rectangular potential with finite width $a$,
\begin{eqnarray}
v_r(x)=V_r\theta\left(-(|x|-\frac{a}{2})\right),
\label{eq:recpotential}
\end{eqnarray}
where $\theta(x)$ is the step function.
The analytic solution of Eq.~(\ref{eq:static}) is obtained as
\begin{eqnarray}
\psi_0(x) =\left\{
\begin{array}{ll}
\tanh\left(\left|x-\frac{a}{2}\right|+\tanh^{-1}\gamma\right) & \left(|x|>\frac{a}{2}\right) \\
\beta/{{\rm cn}\left(\sqrt{K^2+\beta^2} x, \;\frac{K}{\sqrt{K^2+\beta^2}} \right)}    & \left(|x|\leq\frac{a}{2}\right)
  \end{array} \right.,
\end{eqnarray}
if $\beta^2+2(V_r-1)\equiv K^2>0 $ and  
\begin{eqnarray}
\psi_0(x) =\left\{
\begin{array}{ll}
\tanh\left(\left|x-\frac{a}{2}\right|+\tanh^{-1}\gamma\right) & \left(|x|>\frac{a}{2}\right) \\
\psi_0= \beta/{{\rm cd}\left(\beta x,\; \frac{\kappa}{\beta}\right)} & \left(|x|\leq\frac{a}{2}\right)
  \end{array} \right. ,
\end{eqnarray}
if $\beta^2+2(V_r-1)\equiv -\kappa^2<0 $. Here ${\rm cn}(x)$ and ${\rm cd}(x)$ denote the Jacobi elliptic functions; $\beta\equiv\psi_0(0)$ and $\gamma\equiv\psi_0(a/2)$ are determined numerically.
We employ the finite-element method~\cite{zienkiewicz-00} to numerically solve Eq.~(\ref{eq:T(x)}). Details of the finite-element method are given in the Appendix.
\begin{figure}[t]
\begin{center}
\includegraphics[width=\linewidth]{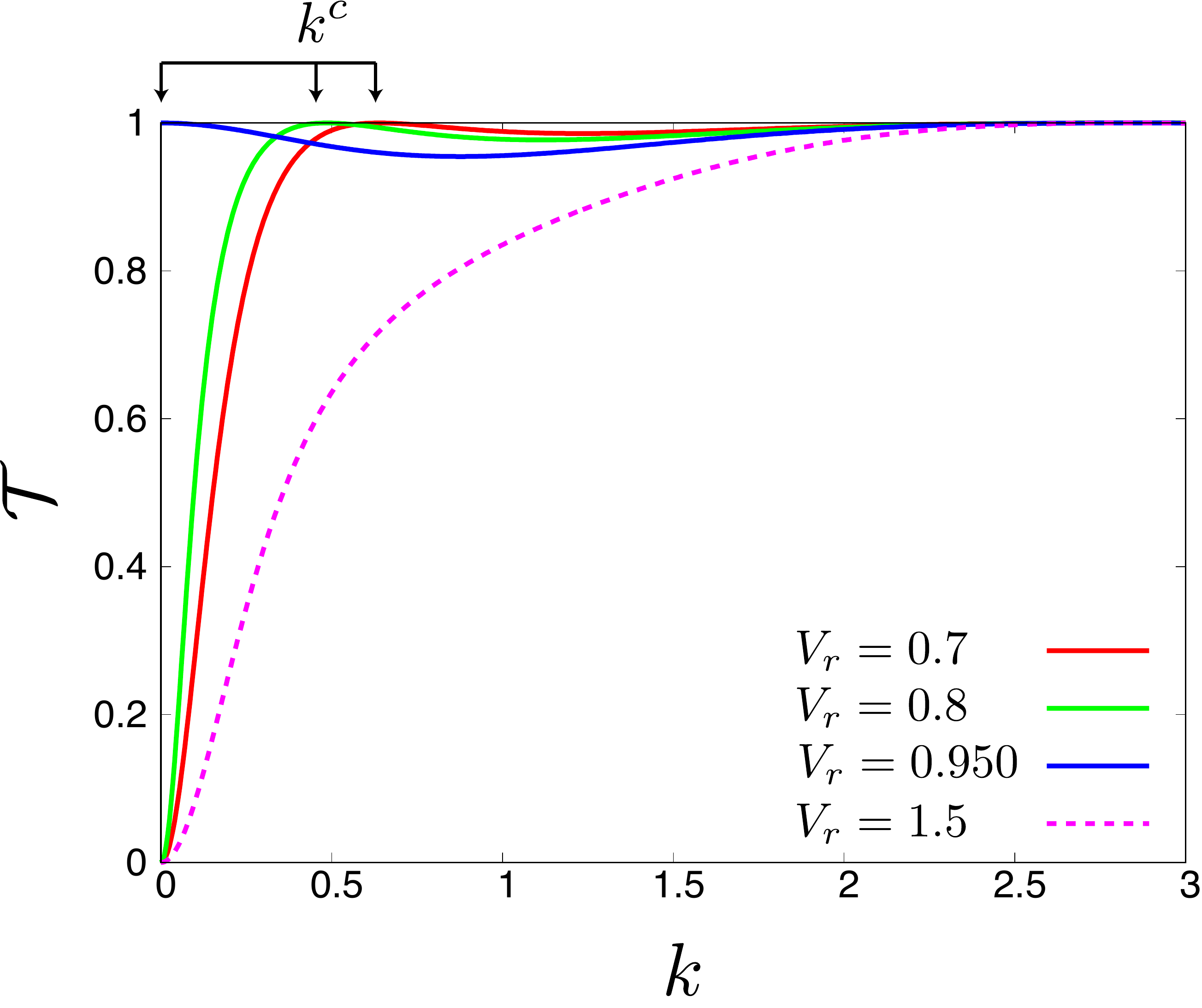}
\caption{Transmission probability of Higgs modes $\mathcal{T}$ as a function of $k$ for a rectangular potential barrier with various strength $V_r$. We set, in the dimensionless unit, $a=1$, $-200\leq x\leq200$, and $N=8000$. The horizontal axis is in the units of $\xi^{-1}$.}
\label{fig:FEM}
\end{center}
\end{figure}

Figure~\ref{fig:FEM} shows the transmission probability of Higgs modes as a function of $k$ for $a=1$. It exhibits qualitatively the same feature as Fig.~\ref{fig:delta}: The perfect transmission occurs at $k^c$ when the strength of the potential is smaller than the critical value $V_r\leq V_r^{\rm c}=0.950$; $k^{\rm c}$ decreases as $V_r$ increases and the perfect transmission no longer occurs when $V_r>V_r^{\rm c}$.
\par
\begin{figure}
\begin{center}
\includegraphics[width=\linewidth]{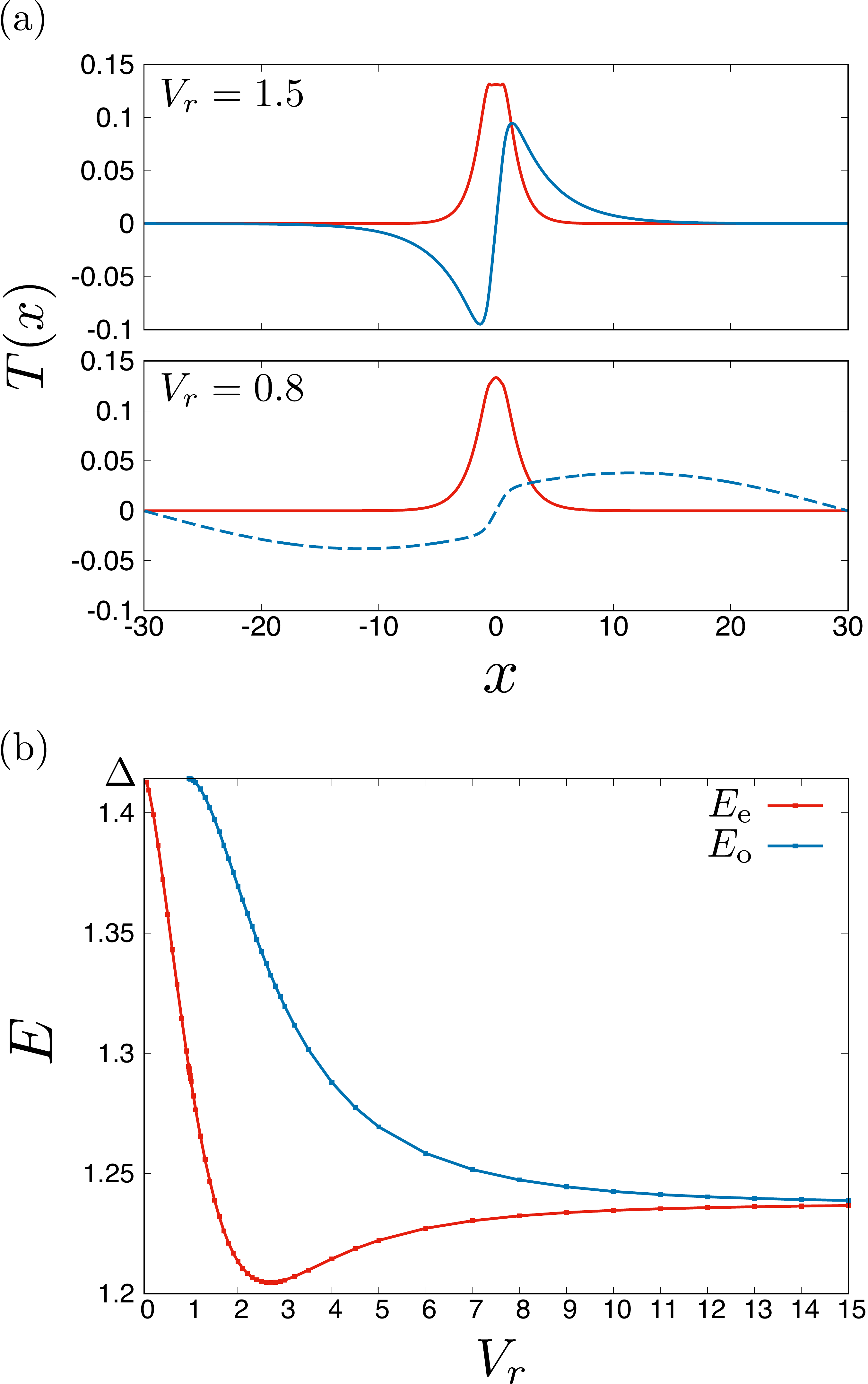}
\caption{(a) Wave functions for the lowest even bound state and the second lowest odd (anti)bound state. The solid lines represent the true bound states and the dashed line represents the antibound state. The odd bound state in the upper panel for $V_r=1.5>V_r^{\rm c}$ becomes an antibound state in the lower panel for $V_r=0.8<V_r^{\rm c}$. We set, in dimensionless unit, $a=1$, $-30a\leq x\leq 30 a$ and $N=1200$. The vertical and horizontal axes are in units of $\sqrt{\left|r_0\right|/u_0}$ and $\xi$, respectively.
(b) Excitation energy of the lowest even bound state ($E_{\rm e}$) and second lowest odd bound state ($E_{\rm o}$) as functions of the potential strength $V_r$. We set, in dimensionless units, $a=1$, $-200a\leq x\leq 200 a$, and $N=8000$. The vertical and horizontal axes are in units of $\sqrt{\left|r_0\right|/W_0}$ and $\left|r_0\right|\xi$, respectively.}
\label{fig:bound_mix}
\end{center}

\end{figure}

In order to study the (anti)bound states, we numerically diagonalize Eq.~(\ref{eq:T(x)}) by the central difference method \cite{cntmethod}. Figure~\ref{fig:bound_mix}(a) shows the wave functions of the (anti)bound states. The excitation energies of the true bound states are plotted as functions of $V_r$ in Fig.~\ref{fig:bound_mix}(b). The lowest odd bound state turns into an antibound state when $V_r\leq V_r^{\rm c}$, as expected. The wave function of the antibound state is delocalized over the system, as shown in the lower panel of Fig.~\ref{fig:bound_mix}(a). Thus, the perfect transmission in Fig.~\ref{fig:FEM} is considered to occur in the same mechanism as the one for a $\delta$-function potential.
\par
\begin{figure}
\begin{center}
\includegraphics[width=\linewidth]{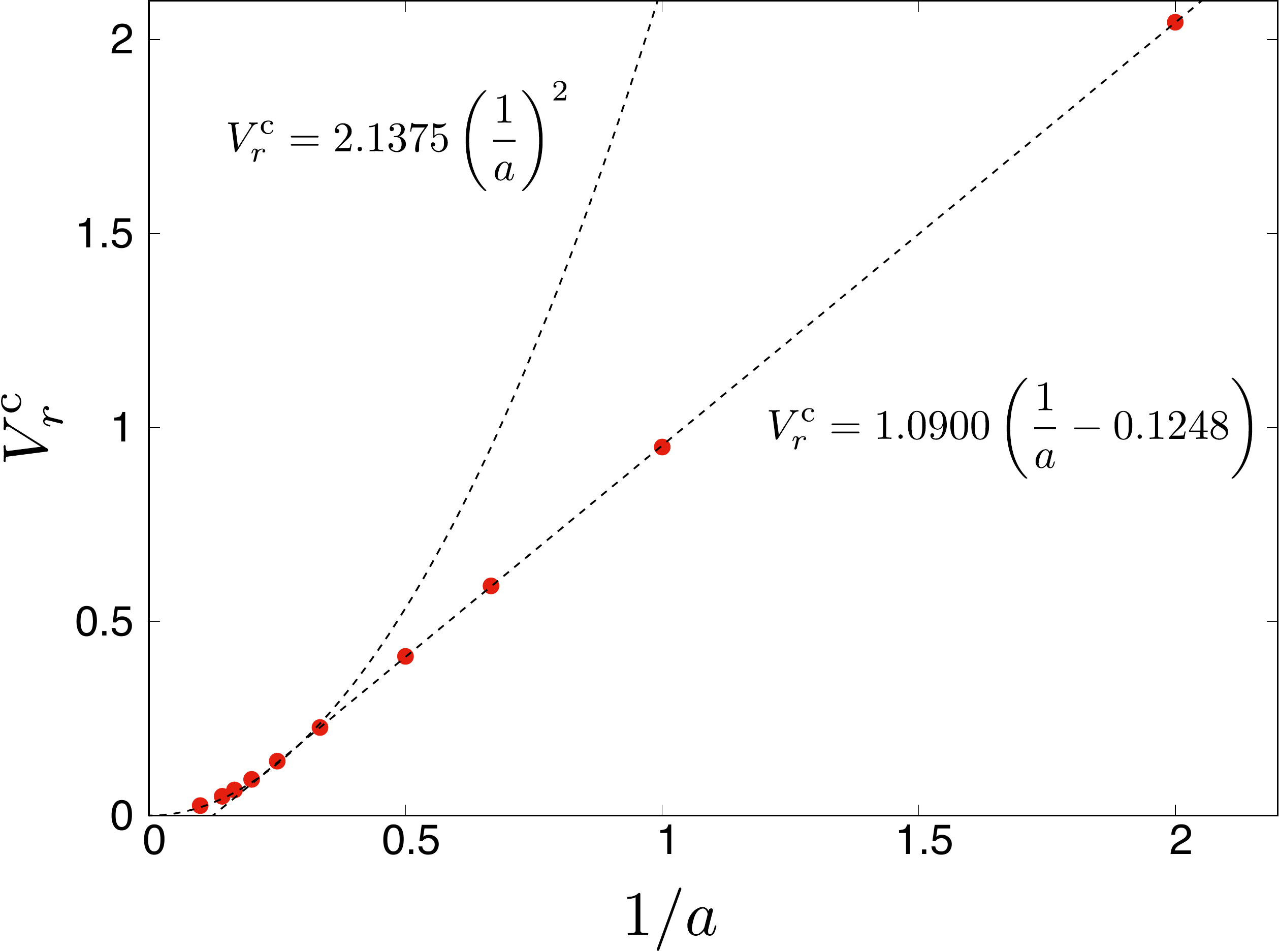}
\caption{Critical strength of the square potential $V_r^{\rm c}$ as a function of the inverse of the barrier width $1/a$.
The vertical and horizontal axes are in units of $\xi^{-1}$ and $\sqrt{\left|r_0\right|/u_0}$, respectively.}
\label{fig:critical_Vr}
\end{center}
\end{figure}

We note that the critical strength of the square potential $V_r^{\rm c}$ depends on the potential width $a$. We calculate $V_r^{\rm c}$ as a function of $a$ and find that $V_r^{\rm c}$ monotonically increases with $1/a$ increasing (the potential narrowing) as shown in Fig.~\ref{fig:critical_Vr}. This implies that a narrow potential barrier is favorable for experimental observation of the perfect transmission of the Higgs mode as well as the transition between the reflectionless and the reflection regime. Figure~\ref{fig:critical_Vr} shows that $V_r^{\rm c}$ quadratically increases with $1/a$ increasing for a wide potential barrier ($a\gtrsim 2\xi$), while $V_r^{\rm c}$ linearly increases with $1/a$ increasing for a narrow potential barrier ($a\lesssim 2\xi$).


\section{Conclusion}\label{sec:Conclusion}
We have studied tunneling properties of Higgs modes in Bose gases in optical lattices through a potential barrier introduced by local modulation of hopping amplitude.
Higgs modes have been found to perfectly transmit through a potential barrier if the barrier strength is weak.
We have found that the perfect transmission disappears at the critical barrier strength above which one of the odd antibound state turns into a true bound state. We demonstrated that the perfect transmission involves resonance with the antibound states.
\par

We propose detection of the perfect transmission of the Higgs mode by Bragg scattering~\cite{bissbort-11}: Exciting the Higgs mode by Bragg laser beams in the presence of a potential barrier introduced by an additional lattice potential in the Gaussian profile or a digital micro-mirror device, one can measure the amplitude of the transmitted wave of the Higgs mode through the potential, from which the transmission probability can be estimated. Observing the perfect transmission of the Higgs mode provides strong evidence for the existence of the antibound states of the Higgs mode.

\par
We finally note that another approach for studying transmission properties of the Higgs mode is the Gutzwiller approximation, which allows us to explore a broader parameter region than the TDGL theory \cite{kovrizhin-07}. The TDGL theory and the Gutzwiller approximation, however, agree quantitatively in the vicinity of the SF-MI transition. Danshita and Tsuchiya compared the two methods regarding the Higgs bound states in Ref.~\cite{danshita-17} and it has been shown, in fact, that the results of the two methods agree well if the system is close enough to the critical point (see, for example, Figs.~8 and 9 in Ref.~\cite{danshita-17}). This demonstrates that these two approximations take into account fluctuations to the same extent in the vicinity of the SF-MI transition point.
Thus, the transmission property of the Higgs mode does not change qualitatively upon approaching the SF-MI transition within the TDGL theory as well as the Gutzwiller approach. The only quantitative changes appear through the scaling of the parameters in Eq.~(\ref{eq:Dless}).
\par
It would be interesting to study the transmission of the Higgs mode in the region where the system is so close to the transition point that the fluctuation of the order parameter gets larger than the order parameter itself. In this regime, the TDGL theory and the Gutzwiller approximation fail to describe the system and one needs an alternative approach based on, for instance, a renormalization-group study. However, that study is beyond the scope of the present paper.


\section*{Acknowledgments}
We acknowledge I. Danshita and T. Nikuni for helpful discussions. In particular, we appreciate N. Hatano for valuable discussions and informing us Ref.~\cite{klaiman-10}. T. N. thanks M. Imada and Y. Yamaji for useful comments.
T. N. was supported by JSPS through Program for Leading Graduate Schools (MERIT). S. T. was supported by Chuo University Grant for Special Research. This work was supported by KAKENHI Grant No.~JP19K03691. 

\vspace{5mm}
\appendix
\section{Finite Element Method}

Equation~(\ref{eq:T(x)}) with $\omega=E$ can be obtained from the variational principle $\delta\mathcal{L}=0$, where the Lagrangian $\mathcal{L}$ is given by
\begin{eqnarray}
\mathcal{L} = \int \:dx \left[\frac{1}{2}\frac{dT^*}{dx}\frac{dT}{dx}+T^*(U(x)-E^2)T^* \right],
\label{eq:lagrangian}
\end{eqnarray}
and $U(x)\equiv 3\psi_0^2(x)-1+v_r(x)$.
\par
We discretize $x$ into $N$ sites, $x_i$ $(i = 1, 2, . . . , N)$, which are referred to as nodes in the literature of the finite element method \cite{zienkiewicz-00}.
We then assign the interpolation function $N_i(x)$ at each $x_i$, which equals unity at $x$ = $x_i$ 
and linearly decreases to zero at the adjacent nodes $x_{i-1}$ and $x_{i+1}$. Namely, the interpolation function is given by
\begin{eqnarray}
N_i(x) \equiv  \left\{ \begin{array}{ll}
    \frac{x-x_{i-1}}{x_i-x_{i-1}} & (x_{i-1}\leq x \leq x_i) \\
    -\frac{x-x_{i+1}}{x_{i+1}-x_i} & (x_{i}\leq x \leq x_{i+1})\\
    0 & ({\rm otherwise})
  \end{array} \right. .
  \label{eq.interpolfunc}
\end{eqnarray}
The function $T(x)$ can be approximated using $N_i(x)$ as 
\begin{eqnarray}
T(x) = \sum_i T_i N_i(x). \label{eq:FEM_T}
\end{eqnarray}
Substituting Eq.~(\ref{eq:FEM_T}) into Eq.~(\ref{eq:lagrangian}), we obtain
\begin{eqnarray}
\mathcal{L} =\bm{T}^\dag\left(\frac{1}{2}\bm{K} +\bm{M}\right) \bm{T},
\label{eq:lagrangian2}
\end{eqnarray}
where $\left(\bm{T}\right)_i\equiv T_i$ and the matrix elements $K_{ij}\equiv \left(\bm{K}\right)_{ij}$ and $M_{ij}\equiv \left(\bm{M}\right)_{ij}$ are given by
\begin{eqnarray}
K_{ij}&=& \int ^{x_{j+1}}_{x_{j-1}} \:dx \frac{dN_i(x)}{dx}\frac{dN_j(x)}{dx},\\
M_{ij}&=& \sum_k \int ^{x_{j+1}}_{x_{j-1}} \:dx N_i(x) N_j(x) N_k(x) .
\end{eqnarray}
The equations for $\bm{T}$ are obtained from the condition $\delta\mathcal{L}/\delta\bm{T}^\dag=0$ as
\begin{eqnarray}
\left(\frac{1}{2}\bm{K} +\bm{M}\right) \bm{T} =\bm{0}.
\end{eqnarray}

In solving the tunneling problem in the presence of a rectangular potential, we find solutions that have the following asymptotic form:
\begin{eqnarray}
T_i \rightarrow \left\{ \begin{array}{l}
T_{1}= e^{-ikh}+r_{\rm h} e^{ikh} \\
T_2 = 1+r_{\rm h} \\
T_{N-1}=t_{\rm h} \\
T_N = t_{\rm h}  e^{ikh} 
\end{array}
 \right. .
\end{eqnarray}
The transmission probability of Higgs modes is given by ${\mathcal T}=\left|t_{\rm h}\right|^2$.

\end{document}